\documentstyle[preprint,epsfig,aps]{revtex}
\tightenlines
\begin{document}
\draft

\title{Thermonuclear Reaction Rate of $^{23}$Mg(p,$\gamma$)$^{24}$Al}

\author{H.~Herndl$^{1}$, M.~Fantini$^{2}$, 
C. ~Iliadis$^{3,4}$\footnote{Corresponding author. 
E-mail: Iliadis@tunl.duke.edu}, P.M.~Endt$^{5}$, and H.~Oberhummer$^{1}$}

\address{$^{1}$ Institut f\"ur Kernphysik, Technische Universit\"at Wien,
	Wiedner Hauptstra{\ss}e 8--10, A--1040 Wien, Austria}
\address{$^{2}$ Dipartimento di Fisica, Universita di Trento, I-38050 Povo 
        (Trento), Italy}
\address{$^{3}$ The University of North Carolina at Chapel Hill, Chapel Hill, 
	      North Carolina 27599-3255, USA}
\address{$^{4}$ Triangle Universities Nuclear Laboratory, Durham, 
         North Carolina 27708-0308, USA}
\address{$^{5}$ R.J. Van de Graaff Laboratorium, Universiteit Utrecht, P.O. 
         Box 80000, 3508 TA Utrecht, The Netherlands}

\date{\today}

\maketitle

\begin{abstract}
Updated stellar rates for the reaction $^{23}$Mg(p,$\gamma$)$^{24}$Al are
calculated by using all available experimental information on $^{24}$Al
excitation energies. Proton and $\gamma$-ray partial widths for
astrophysically important resonances are derived from 
shell model calculations. 
Correspondences of experimentally observed $^{24}$Al levels
with shell model states are based on application of the 
isobaric multiplet mass equation. Our new rates suggest that the 
 $^{23}$Mg(p,$\gamma$)$^{24}$Al reaction influences the nucleosynthesis
in the mass A$>$20 region during thermonuclear runaways on massive white
dwarfs.
\end{abstract}


\narrowtext

\section{Introduction}
Explosive stellar burning of hydrogen in the mass A$>$20 range
is characterized by a large number of proton capture
reactions and $\beta$-decays. The resulting network of nuclear processes
is called the rp--process \cite{wal81}. This process might be responsible for
the energy production and nucleosynthesis in a variety of astrophysical sites
with different temperature and density conditions. For example,
in novae, typical peak temperatures range from
$T_9 = 0.2 - 0.4$ \cite{POL}, with $T_9$ the temperature in GK. For x--ray
bursts and accreting black holes the rp--process could take place
at very high temperatures in excess of $T_{9}$=1 \cite{TAA,JIN}.
Stellar rates of several proton capture reactions relevant for the
rp--process were estimated by Wallace and Woosley \cite{wal81} and by
Wiescher et al. \cite{wie86}.
Recently, some of these rates were updated with new experimental 
information and improved theoretical models \cite{van94,her95,sch97,ili98}.

At low stellar temperatures T$_{9}$$<$0.1 the isotope $^{23}$Mg is 
synthesized in the NeNa-cycle. Under such conditions the $\beta$-decay of
$^{23}$Mg (T$_{1/2}$=11.3 s) and the subsequent $^{23}$Na(p,$\alpha$) reaction 
convert material back into $^{20}$Ne, giving rise to cycling of material in the
NeNa mass range. If the stellar temperature is 
sufficiently high the proton capture reaction on $^{23}$Mg becomes faster than
the competing $\beta$-decay. In this case the reaction flow breaks out of the
NeNa mass region and a whole range of heavier nuclei could be synthesized, 
depending on the temperature-density conditions and the duration of the 
astrophysical event. This scenario, for example, might be responsible for the
synthesis of elements such as Si, S and Ar, which have been found to be
overabundant in the ejecta of ONeMg novae \cite{POL}.  
Therefore, a quantitative estimate of the stellar reaction rate for
$^{23}$Mg(p,$\gamma$)$^{24}$Al is important in order to model the
nucleosynthesis in the mass A$>$20 range. 
At very high temperatures above T$_{9}$=1 the $^{23}$Mg(p,$\gamma$)$^{24}$Al 
reaction is of minor importance, since the isotope $^{23}$Mg
is bypassed via the sequence $^{21}$Na(p,$\gamma$)$^{22}$Mg(p,
$\gamma$)$^{23}$Al(p,$\gamma$)$^{24}$Si($\beta^+ \nu$)$^{24}$Al.

The reaction rate for $^{23}$Mg(p,$\gamma$)$^{24}$Al was previously
estimated by Wallace and Woosley \cite{wal81}, and their
calculation was based on a single resonance only. Subsequently the reaction 
rate calculation was improved by Wiescher et al.~\cite{wie86} who considered three 
resonances and in addition a contribution from the direct capture 
process. The most recent estimate was published by Kubono et al.~\cite{kub95}.
These authors, who investigated the level structure of $^{24}$Al near the proton
threshold by using the $^{24}$Mg($^{3}$He,t)$^{24}$Al charge-exchange reaction,
based the reaction rate estimate on their experimentally determined $^{24}$Al 
excitation energies and spin-parity restrictions.

We present a reanalysis of the $^{23}$Mg(p,$\gamma$)$^{24}$Al reaction rate
for several reasons. First, a recent experimental study of the 
$^{24}$Mg($^{3}$He,t)$^{24}$Al charge-exchange reaction was published by
Greenfield et al. \cite{gre91}. The excitation energies of proton threshold
levels in $^{24}$Al measured by the two groups differ by about 30--50 keV. This
difference might change the resulting reaction rates appreciably. Second,
the authors of Ref. \cite{kub95} conclude that the analog assignments of the
two lowest-lying proton threshold states in $^{24}$Al are still uncertain,
resulting in large errors of the derived stellar reaction rates.
Third, the proton and $\gamma$-ray partial widths of the resonances in
question have never been measured. These quantities were crudely estimated in 
Ref. \cite{kub95} by adopting `typical' single-particle 
spectroscopic
factors and `average' $\gamma$-ray transition strenghts from Ref. \cite{wie86}.

In this work we use $^{24}$Al excitation energies recommended by Ref. \cite{end98} 
which are based on previously published experimental results.
We present additional support for the analog assignments of the 
proton threshold levels in $^{24}$Al. Furthermore, we calculate the proton and
$\gamma$-ray partial widths of astrophysically important levels by using the
nuclear shell model.
In Sect. II we describe briefly the formalism for calculating stellar 
reaction rates. Experimental and theoretical nuclear input parameters are 
presented in Sect. III. Results and astrophysical implications are discussed
in Sect. IV. A summary is given in Sect. V.

\section{Calculation of Stellar Reaction Rates}

The proton capture cross sections on sd-shell nuclei are predominantly 
determined
by summing the contributions from isolated resonances corresponding to 
unbound  compound nuclear states and from the nonresonant direct capture (DC) 
process. In the following we briefly describe the method of calculating the 
resonant and nonresonant (DC) contributions to the stellar reaction rates.

\subsection{Resonant Reaction Contributions}

For the reaction under consideration here the resonances are narrow and 
isolated. The resonant rate contribution can be calculated from resonance
energies E$_{i}$ and resonance strengths $\omega\gamma_{i}$ (both in units 
of MeV)
\cite{rol88}
\begin{equation}
\label{eq-rrate}
N_{\rm A} <\sigma v>_{\rm r} = 1.54 \times 10^{11} (\mu T{_9})^{-3/2}
\sum_i (\omega \gamma)_i \exp{(-11.605 E_i / T_9)} \hspace{5mm}
{\rm cm}^3 {\rm mole}^{-1} {\rm s}^{-1} \quad .
\end{equation}
The resonance strength $\omega\gamma$ for a (p,$\gamma$) reaction is given by
\begin{equation}
\omega \gamma = \frac {2J+1}{2(2j_t+1)} \frac{\Gamma_{\rm p}
\Gamma_{\rm\gamma}}{\Gamma_{\rm tot}} \quad ,
\end{equation}
where $J$ and $j_{\rm t}$ are the spins of the resonance and the
target nucleus, respectively, and the total width $\Gamma_{\rm tot}$
is the sum of the proton partial width $\Gamma_{\rm p}$ and the $\gamma$-ray 
partial width $\Gamma_{\gamma}$.

The proton partial width $\Gamma_{\rm p}$ can be estimated from the
single--particle spectroscopic factor S and the single--particle width
$\Gamma_{\rm sp}$ of the resonance by using ~\cite{sch63}
\begin{equation}
\label{SF}
\Gamma_{\rm p} = C^2 S\cdot\Gamma_{\rm sp} \quad,
\end{equation}
where $C$ is the isospin Clebsch--Gordan coefficient. Spectroscopic
factors S are calculated in this work by using the nuclear shell model
(Sect.~III). Single--particle widths $\Gamma_{\rm sp}$ are obtained 
from resonant scattering phase shifts generated by an appropriate folding 
potential (see below). In this context, the partial width $\Gamma_{\rm sp}$ is 
defined as the energy interval over which the resonant 
phase shift varies from $\pi / 4$ to $3\pi / 4$. 

Gamma--ray partial widths for specific electromagnetic transitions are 
expressed in terms of reduced transition probabilities 
B($J_{i}\rightarrow J_{f}$;L) which contain the nuclear structure information 
of the states involved in the transition \cite{bru77}. In the present work, 
the reduced transition probabilities are calculated in the framework of
the nuclear shell model (Sect. III). The total $\gamma$--ray width 
$\Gamma_{\gamma}$ of a particular resonance is given by the sum over  
partial $\gamma$--ray widths for transitions to all possible lower--lying 
nuclear states.

\subsection{Nonresonant Reaction Contributions}

The nonresonant proton capture cross section is calculated by using the 
direct capture (DC) model described in\,\cite{kim87,obe91,moh93}. 
The DC cross section $\sigma^{\rm DC}_i$ for a particular
transition is determined by the overlap of the scattering wave function
in the entrance channel, the bound--state wave function in the exit channel 
and the electromagnetic multipole transition operator. Usually, only 
the dominant E1 transitions have to be taken into account. Wave functions 
are obtained by using a real folding potential given by \cite{obe91,kob84}
\begin{equation}
\label{eq-fold}
V(R) = 
  \lambda\,V_{\rm F}(R) 
  = 
  \lambda\,\int\int \rho_a({\bf r}_1)\rho_A({\bf r}_2)\,
  v_{\rm eff}\,(E,\rho_a,\rho_A,s)\,{\rm d}{\bf r}_1{\rm d}{\bf r}_2 \quad .
\end{equation}
Here $\lambda$ represents a potential strength parameter close to unity, and 
$s = |{\bf R} + {\bf r}_2 - {\bf r}_1|$, with $R$ the separation of the 
centers-of-mass of the projectile and the target nucleus. The mass density 
distributions $\rho_a$ and $\rho_A$ are either derived from measured charge 
distributions\,\cite{vri87} or calculated by using structure models (e.g.,
Hartree--Fock calculations). For the effective nucleon--nucleon interaction 
$v_{\rm eff}$ we used the DDM3Y parametrization\,\cite{kob84}. The imaginary 
part of the potential has been neglected due to the small flux into other 
reaction channels.

The total nonresonant cross section $\sigma_{\rm nr}$ is determined by  
summing contributions of direct capture transitions to all 
bound states with single-particle spectroscopic factors S$_{i}$: 
\begin{equation}
\label{NR}
\sigma_{\rm nr} = \sum_{i} \: (C^{2} S)_i\sigma^{\rm DC}_i \quad .
\end{equation}
The astrophysical S--factor of a charged--particle induced reaction is defined 
by \cite{rol88}
\begin{equation}
S (E) = E \exp{(2 \pi \eta)} \sigma (E) ,
\end{equation}
with $\eta$ denoting the Sommerfeld parameter.
If the S--factor depends only weakly on the bombarding energy the nonresonant 
reaction rate as a function of temperature T$_{9}$ can be expressed as
\begin{eqnarray}
\label{eq-nrate}
N_{\rm A} <\sigma v>_{\rm nr} & = & 7.833 \times 10^9  
\left( \frac{Z_1 Z_2}{A T_9^2} \right) ^{1/3} S(E_0) [{\rm MeV barn}]
\nonumber\\
 & \times & \exp{ \left[ -4.249 \left( \frac{Z_1^2 Z_2^2 A}{T_9}
\right) ^{1/3} \right] } \hspace{1cm}
{\rm cm}^3 {\rm mole}^{-1} {\rm s}^{-1} \quad ,
\end{eqnarray}
with Z$_{1}$ and Z$_{2}$ the charges of the projectile and target, respectively, 
and $A$ the reduced mass (in amu). The quantity E$_{0}$ denotes the 
position of the Gamow peak corresponding to the effective bombarding energy 
range of stellar burning.

\section{Experimental and Theoretical Input Parameters}

In this section we present a discussion of excitation energies, spectroscopic
factors and $\gamma$--ray partial widths which enter in the calculation of 
stellar reaction rates.

Experimental excitation energies in $^{24}$Al below E$_{x}$=3 MeV have been 
compiled by Endt
\cite{end90}. The energies listed in Table 24.23 of Ref.~\cite{end90} 
are based on 
$^{24}$Mg($^{3}$He,t) and (p,n) reaction studies performed prior to 1990.
Three recent charge--exchange reaction studies \cite{kub95,gre91,kia89} 
also report $^{24}$Al excitation energies. In the present work we use the  
E$_{x}$($^{24}$Al) values compiled and evaluated by Ref. \cite{end98} which are based on 
experimental information. The values
measured in the (p,n) study of Kiang et al. \cite{kia89} have been disregarded
because of the superior energy resolution and counting statistics of the ($^{3}$He,t) reaction
studies. Our adopted 
excitation energies (Tables I and II) differ from the results of Kubono et al. 
\cite{kub95} on average by about 20 keV. 

Spectroscopic factors and reduced $\gamma$--ray transition strengths for the
levels of astrophysical interest are calculated by using the nuclear 
shell model. This procedure requires the identification of experimentally 
observed $^{24}$Al levels with calculated shell model states. For bound 
$^{24}$Al states there is a one--to--one correspondence. However, the spins and
parities of experimentally observed unbound states are not known uniquely, 
resulting in ambiguities for the shell model assignments. Level assignments 
based on a comparison of experimental and shell model excitation energies 
alone are not useful either, mainly due to the difficulty of the shell model 
in producing accurate E$_{x}$ values. In this work we have used a method 
described in Ref. \cite{ili98} to which the reader is referred for 
details. In brief, experimental excitation 
energies of $^{24}$Na and $^{24}$Mg states,
for which the spins and parities are well--known, are used together with
the isobaric multiplet mass equation (IMME) \cite{orm89} in order to calculate
excitation energies of $^{24}$Al analog states:
\begin{equation}
\label{eq-IMME}
E_{\rm x}(^{24}{\rm Al})=2E_{\rm x}(^{24}{\rm Mg})-E_{\rm x}(^{24}{\rm Na})+
2 [ c-c({\rm g.s.}) ]
\quad .
\end{equation}
The coefficient c is a measure for the isotensor Coulomb energy of a
specific isobaric triplet and is estimated in this work by using the nuclear
shell model (see below). Correspondences between experimentally
observed $^{24}$Al levels and shell model states are found by a) minimizing 
the difference between measured excitation energies and E$_{x}$ values 
calculated from Eq. (\ref{eq-IMME}), and b) matching experimentally 
determined spin-parity
restrictions with shell model quantum numbers. 
It can be seen from our results listed in Table \ref{tab-levels} 
that the experimentally observed $^{24}$Al states at E$_{x}$=2349 and 2534 keV
most likely correspond to the 3$^{+}_{3}$ and 4$^{+}_{2}$ 
shell model states, respectively, in agreement with the tentative assignments 
of Ref.~\cite{kub95}. However, the experimental states at E$_{x}$=2810 and 
2900 keV most likely correspond to the 2$^{+}_{4}$ and 3$^{+}_{4}$ 
shell model states, respectively, in contradiction with the results of 
Kubono et al. \cite{kub95}.

Shell model calculations have been performed by using the code
OXBASH \cite{bro84}. The isospin--nonconserving (INC)
interaction of Ormand and Brown \cite{orm89} is employed for the calculation 
of wave functions and excitation energies of T=1 triplet states in the mass
A=24 system. Coefficients c in Eq. (8) are estimated from theoretical 
excitation energies. Shell model wave functions are used for calculating 
single--particle spectroscopic factors and reduced transition probabilities 
for M1-- and E2-- $\gamma$-ray decays. 
Resulting values of the resonance parameters are listed in Table
\ref{tab-res}. It can be seen that the $\gamma$-ray partial width is much 
smaller than the proton partial width for all resonances considered.
Therefore, the resonance strength $\omega\gamma$ depends mainly on the
value of $\Gamma_{\gamma}$.
Table \ref{tab-gg} displays the calculated $\Gamma_{\gamma}$ values 
for the $^{24}$Al unbound states of main interest in this work together with
theoretical and experimental $\Gamma_{\gamma}$ values of the corresponding
$^{24}$Na mirror states. It can be seen that in the case of $^{24}$Na  
the shell model $\gamma$-ray widths  are in excellent agreement with the 
experimental values deduced from lifetime measurements.

In Table II our resonance strengths are compared to previous results
\cite{kub95}. For the first resonance the strengths are very similar, since
the experimentally measured lifetime of the $^{24}$Na mirror state 
was used in Ref. \cite{kub95}. For the second resonance our strength 
$\omega\gamma$ is considerably smaller than the value of Kubono et al.
\cite{kub95} who adopted an `average' $\gamma$-ray transition strength from
Ref. \cite{wie86}. For the third and fourth resonance the spin--parity
assignments are interchanged compared to Ref. \cite{kub95} (see above)
which explains the discrepancy of the $\omega\gamma$ values. It should be
noted that the latter two resonances have a negligible influence on 
the stellar reaction rates (Sect. IV).

The parameters for the direct capture (DC) contribution to the stellar 
reaction rates
are presented in Table \ref{tab-dc}. All transitions considered 
are displayed together with our calculated shell model spectroscopic
factors. With this information, the total astrophysical S--factor for the 
direct capture process into all bound states has been determined
(Sect. IIb). For bombarding energies below 1 MeV the S--factor can be 
expressed as
\begin{equation}
 S(E) = 22.5 - 1.1 \times 10^{-2} E + 6.9 \times 10^{-6} E^{2}~~{\rm keV 
 \cdot b}.
\end{equation}
Our derived direct capture S-factor is about 10 \% smaller than the 
results of Wiescher et al. \cite{wie86} which were also adopted by Ref. 
\cite{kub95}.

\section{Discussion}

The recommended stellar rates of the $^{23}$Mg(p,$\gamma$)$^{24}$Al reaction can
be parametrized for temperatures below T$_{9}$=2 by the expression \cite{wie86}
\begin{equation}
N_{\rm A} <\sigma v> = \sum_i A_i/T932 \exp{(-B_i/T_9)}
+ C/T923 \exp{(-D/T913)} \quad {\rm cm}^3 
{\rm mole}^{-1} {\rm s}^{-1} \quad ,
\label{eq-rate}
\end{equation}
where, for example, T932 stands for $T_9^{3/2}$. The first and second term 
in Eq. (\ref{eq-rate}) represent the contributions of all narrow resonances and 
the direct capture process, respectively. The parameters $A_i$, $B_i$, $C$ and 
$D$ are listed in Table \ref{tab-para}. 

\begin{figure}
\centerline{\epsfig{figure=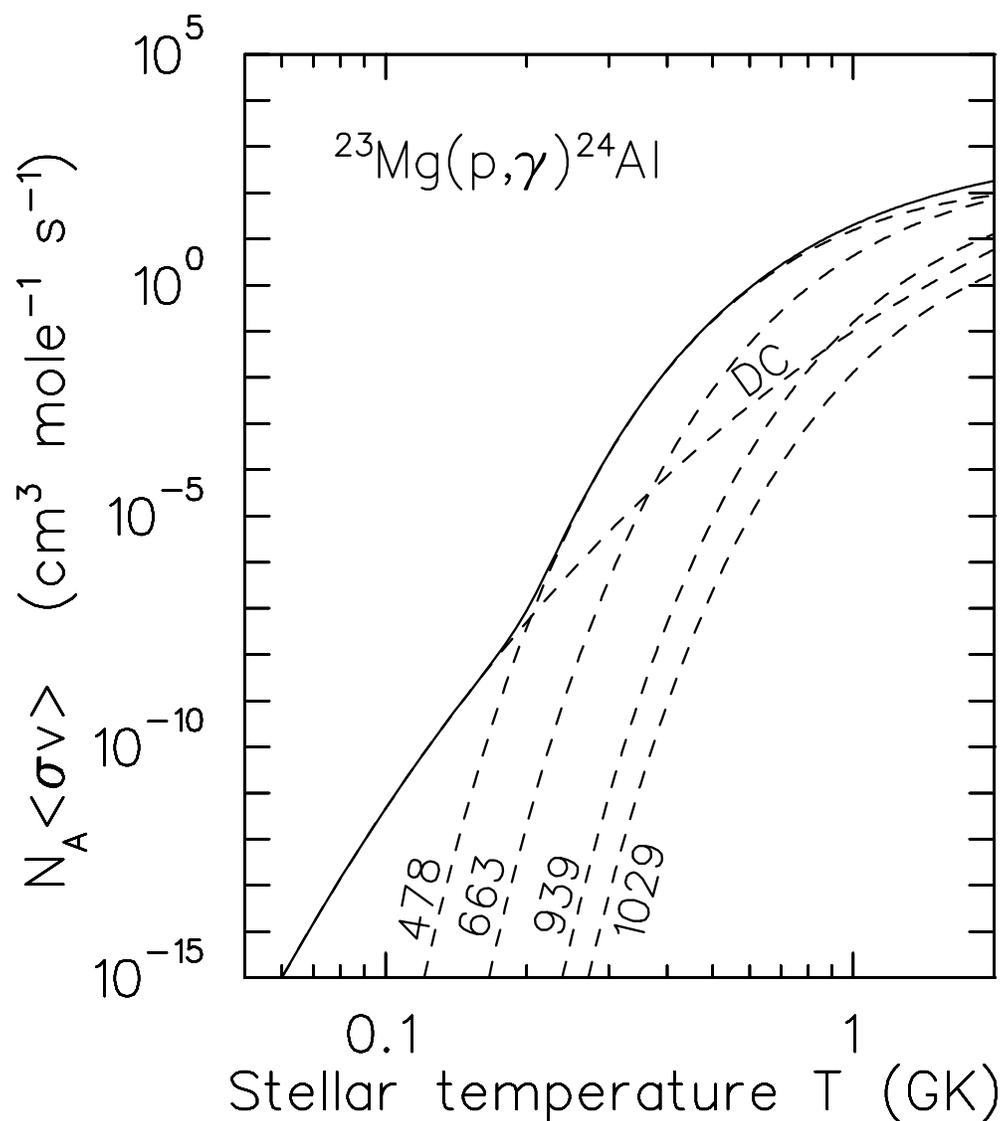,width=13cm}}
\caption[fig1]
{Total stellar rate (solid line) and individual contributions (dashed lines)
for the reaction $^{23}$Mg(p,$\gamma$)$^{24}$Al.}
\label{fig-rate}
\end{figure}

\begin{figure}
\centerline{\epsfig{figure=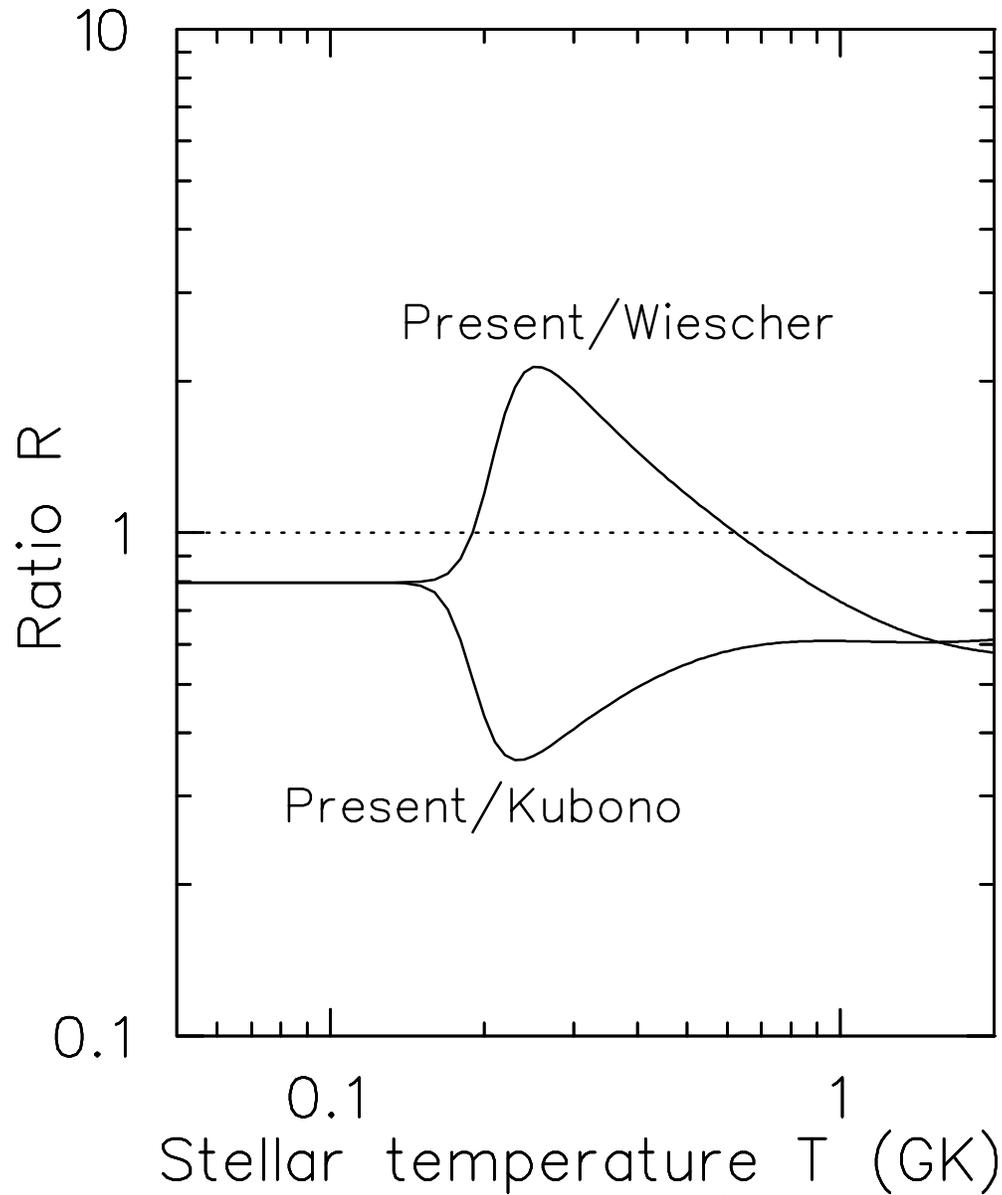,width=13cm}}
\caption[fig2]
{Ratio of the present reaction rate to previous results of Wiescher et al.
\cite{wie86} and Kubono et al. \cite{kub95}. The reaction rates are based on 
measured $^{24}$Al excitation energies (see text) which are most likely Gaussian
distributed.} 
\label{fig-ratio}
\end{figure}

The various contributions to the total reaction rate are displayed in 
Fig.~\ref{fig-rate}. The direct capture process determines the stellar rates
at low temperatures of $T_9 < 0.2$. The E$_{R}$=478 keV resonance 
dominates the reaction rates in the range T$_{9}$=0.2--1.0.
The E$_{R}$=663 keV resonance is of importance at high temperatures above
T$_{9}$=1 only. The resonances at E$_{R}$=939 and 1029 keV are negligible 
over the whole temperature range.
Our results are compared in Fig.~\ref{fig-ratio} with previous work 
\cite{wie86,kub95}. At stellar temperatures above T$_{9}$=1 the reaction
rates of the present work are smaller than the results of Refs. 
\cite{wie86,kub95} by about 70\%. In the temperature range T$_{9}$=0.2-0.5 
important for hydrogen burning in novae (see below) the present reaction rates 
deviate up to a factor
of 3 from the values given in Ref. \cite{kub95}, and up to a factor of 2
from the results of Ref. \cite{wie86}. The reaction rates for
$^{23}$Mg+p are therefore now based on more consistent experimental and theoretical input
parameters. 

Figure 3 presents temperature and density conditions for which the proton
capture reaction on $^{23}$Mg and the $^{23}$Mg $\beta$-decay are of equal
strength. The solid line is calculated by 
assuming a hydrogen mass fraction of X$_{H}$=0.365 \cite{POL}.  
Recent results of hydrodynamic studies of ONeMg novae \cite{POL} are also shown 
in Fig. 3. The full circles represent temperature and density conditions at the 
peak of the thermonuclear runaway for accretion onto white dwarfs of different 
inital masses (1.00M$_{\odot}$, 1.25M$_{\odot}$ and 1.35M$_{\odot}$). 
Our results indicate that for white dwarfs of masses 
$\leq$1.25M$_{\odot}$ the proton-capture reaction on $^{23}$Mg is slower than
the competing $\beta$-decay and, therefore, is of minor importance for the
resulting nucleosynthesis. However, for accretion onto very massive white
dwarfs (1.35M$_{\odot}$ model of Ref. \cite{POL}) the 
$^{23}$Mg(p,$\gamma$)$^{24}$Al reaction dominates over the $^{23}$Mg 
$\beta$-decay and will influence the nucleosynthesis in the mass A$>$20
range. 
Stellar network calculations are underway in order to investigate
quantitatively the implications of our new $^{23}$Mg+p reaction rates and their
corresponding uncertainties. 
The results will be published in a forthcoming paper \cite{ili98b}. 

\section{Summary and conclusions}
Improved estimates of reaction rates for $^{23}$Mg(p,$\gamma$)$^{24}$Al are
presented in this work. Resonance energies are derived from all available
experimental information on $^{24}$Al excitation energies. Proton and
$\gamma$-ray partial widths of astrophysically important resonances are 
estimated from single-particle spectroscopic factors and reduced $\gamma$-ray 
transition probabilities, respectively, derived from shell model calculations.
Correspondences of experimentally observed $^{24}$Al levels with shell model
states are based on application of the isobaric multiplet mass equation.
In the temperature range T$_{9}$=0.2--0.5 important for the nucleosynthesis
in novae
the present reaction rates deviate up to a factor of three from previous results.
Our new stellar rates suggest that the $^{23}$Mg(p,$\gamma$)$^{24}$Al
reaction will influence the nucleosynthesis in the mass A$>$20 region 
during thermonuclear runaways on massive white dwarfs. Quantitative predictions
have to await the results of large-scale stellar network calculations. 

\acknowledgments
We are grateful to C. van der Leun for providing helpful comments. One of us (M.F.)
would like to thank the Institut f\"{u}r Kernphysik for hospitality and the
Instituto Nazionale di Fisica Nucleare for partial support.
This work was supported in part by Fonds zur F\"orderung der wissenschaftlichen
Forschung (FWF project S7307--AST) and by the U.S. Department of Energy under
Contract No. DE-FG02-97ER41041.

\begin{table}
\caption{Comparison of excitation energies (in MeV).}
\begin{center}
\footnotesize
\begin{tabular}{llllllll}
\multicolumn{2}{c}{$^{24}$Al$^{a}$} &
\multicolumn{2}{c}{$^{24}$Na$^{a}$} &
\multicolumn{2}{c}{$^{24}$Al (IMME) $^{b}$} &
\multicolumn{2}{c}{$^{24}$Al (OXBASH) $^{c}$} \\
$E_{x}$ & $J^{\pi}$ & $E_{x}$ & $J^{\pi}$ & $E_{x}$ & $J^{\pi}$ & $E_{x}$ & 
$J^{\pi}$ \\ \hline
 0.0 & $4^+$ & 0.0 & $4^+$ & 0.0 & $4^+$ & 0.0 & $4^+$ \\
 0.426 & $1^+$ & 0.472 & $1^+$ & 0.441 & $1^+$ & 0.448 & $1^+$ \\
 0.510 & $2^+$ & 0.563 & $2^+$ & 0.450 & $2^+$ & 0.580 & $2^+$ \\
 1.107 & $(1-3)^+$ & 1.347 & $1^+$ & 1.073 & $1^+$ & 1.100 & $1^+$ \\
 1.130 & $(1-3)^+$ & 1.341 & $2^+$ & 1.069 & $2^+$ & 1.126 & $2^+$ \\
 1.275 & $3^+$ & 1.345 & $3(^+)$ & 1.283 & $3^+$ & 1.374 & $3^+$ \\
 1.559 & ($5^+$) & 1.512 & $5^+(3^+)$ & 1.552 & $5^+$ & 1.554 & $5^+$ \\
 1.559 & ($2^+$) & 1.846 & $2^+$ & 1.541 & $2^+$ & 1.590 & $2^+$ \\
 1.634 & $3^+$ & 1.886 & $3^+$ & 1.715 & $3^+$ & 1.729 & $3^+$ \\
 2.349 & $(3^+)$ & 2.514 & $3^+$ & 2.305 & $3^+$ & 2.176 & $3^+$ \\
 2.534 & $(4,5)^+$ & 2.563 & $4^+(2^+)$ & 2.476 & $4^+$ & 2.541 & $4^+$ \\
 2.810 & $2^+$ & 2.978 & $2^+(3^+)$ & 2.803 & $2^+$ & 2.837 & $2^+$ \\
 2.900 & $(1-3)^+$ & 2.904 & $3^+$ & 2.737 & $3^+$ & 2.629 & $3^+$ \\
\end{tabular}
\end{center}
$^{a}$ Experimental values evaluated and compiled in Ref. \cite{end98}.\\
$^{b}$ Calculated by using Eq. (8).\\
$^{c}$ Calculated by using the nuclear shell model.\\
\label{tab-levels}
\end{table}

\begin{table}[bh]
\caption[Resonance parameters 1]
{Resonance parameters for the reaction $^{23}$Mg(p,$\gamma$)$^{24}$Al.} 
\begin{center}
\footnotesize
\begin{tabular}{ccccccc}
$E_{\rm x}$ (MeV) $^{a}$ & $J^{\pi}$ & 
$E_{\rm R}^{cm}$ (MeV)  $^{b}$ & $\Gamma_{\rm p}$ (meV) & 
$\Gamma_{\gamma}$ (meV) & $\omega\gamma$ (meV) & $\omega\gamma$ (meV) $^{c}$ \\
\hline
2.349$\pm$0.020 & $3_3^+$ & 0.478 & 185 & 33 & 25 & 27 \\
2.534$\pm$0.013 & $4_2^+$ & 0.663 & $2.5 \times 10^3$ & 53 & 58 & 130 \\
2.810$\pm$0.020 & $2_4^+$ & 0.939 & $9.5 \times 10^5$ & 83 & 52 & 11 \\
2.900$\pm$0.020 & $3_4^+$ & 1.029 & $3.4 \times 10^4$ & 14 & 12 & 16 \\
\end{tabular}
\end{center}
$^{a}$ Experimental values adopted from Ref. \cite{end98}.\\
$^{b}$ Calculated from column 1 and Q$_{p\gamma}$=1871$\pm$4 keV 
       \cite{aud95}.\\
$^{c}$ From Ref. \cite{kub95}.\\
\label{tab-res}
\end{table}

\begin{table}[htb]
\caption{Gamma-ray partial widths (in meV) of $^{24}$Al and $^{24}$Na levels.}
\begin{center}
\footnotesize
\begin{tabular}{cccc}
$J^{\pi}$ & $\Gamma_{\gamma}^{\rm SM}$ ($^{24}$Al) $^{a}$ &
$\Gamma_{\gamma}^{\rm SM}$ ($^{24}$Na) $^{a}$ &
$\Gamma_{\gamma}^{\rm exp}$ ($^{24}$Na) $^{b}$ \\
\hline
$3_3^+$ & 33 & 43 & 46 $\pm$ 14 \\
$4_2^+$ & 53 & 51 & $> 27$ \\
$2_4^+$ & 83 & 93 & $> 27$ \\
$3_4^+$ & 14 & 15 & 13 $\pm$ 2 \\
\end{tabular}
\end{center}
$^{a}$ Shell model values for the $^{24}$Al -- $^{24}$Na mirror pair.\\
$^{b}$ Determined from measured lifetimes of $^{24}$Na levels.\\
\label{tab-gg}
\end{table}

\begin{table}[htb]
\caption{Shell model spectroscopic factors of $^{24}$Al bound states.} 
\begin{center}
\footnotesize
\begin{tabular}{ccccc}
E$_{x}$(MeV) & J$^{\pi}$  & \multicolumn{3}{c}{C$^{2}$S}\\ 
 &  & \multicolumn{1}{c}{p $\rightarrow$ 1d$_{3/2}$} 
 & \multicolumn{1}{c}{p $\rightarrow$ 1d$_{5/2}$} 
 & \multicolumn{1}{c}{p $\rightarrow$ 2s$_{1/2}$}\\ 
\hline
0.000 & 4$^+$ &    --     & 0.39 & -- \\
0.426 & 1$^+$ &  0.0010 & 0.69 & 0.0031 \\
0.510 & 2$^+$ &  0.0000 & 0.29 & 0.052 \\
1.107 & 1$^+$ &  0.098 & 0.016 & 0.36 \\
1.130 & 2$^+$ &  0.084 & 0.0001 & 0.30 \\
1.275 & 3$^+$ &  0.024 & 0.0057 & -- \\
1.559 & 5$^+$ &   --    &   --   & --  \\
1.559 & 2$^+$ &  0.11 & 0.088 & 0.19 \\
1.634 & 3$^+$ &  0.0009 & 0.19 & -- \\ 
\end{tabular}
\end{center}
\label{tab-dc}
\end{table}

\begin{table}[t]
\caption{Recommended parameters for the $^{23}$Mg(p,$\gamma$)$^{24}$Al 
reaction rate.$^{a}$}
\begin{center}
\begin{tabular}{cccc}
$A_i$ & $B_i$ & $C$ & $D$ \\ \hline
$4.02 \times 10^3$ & 5.56  & $3.72 \times 10^8$ & 21.95 \\
$9.59 \times 10^3$ & 7.71  &                    &       \\
$8.52 \times 10^3$ & 10.91 &                    &       \\
$2.0  \times 10^3$ & 11.95 &                    &       \\
\end{tabular}
\end{center}
$^{a}$ The total stellar reaction rate is given by Eq. (10).\\
\label{tab-para}
\end{table}
\newpage
\begin{figure}
\centerline{\epsfig{figure=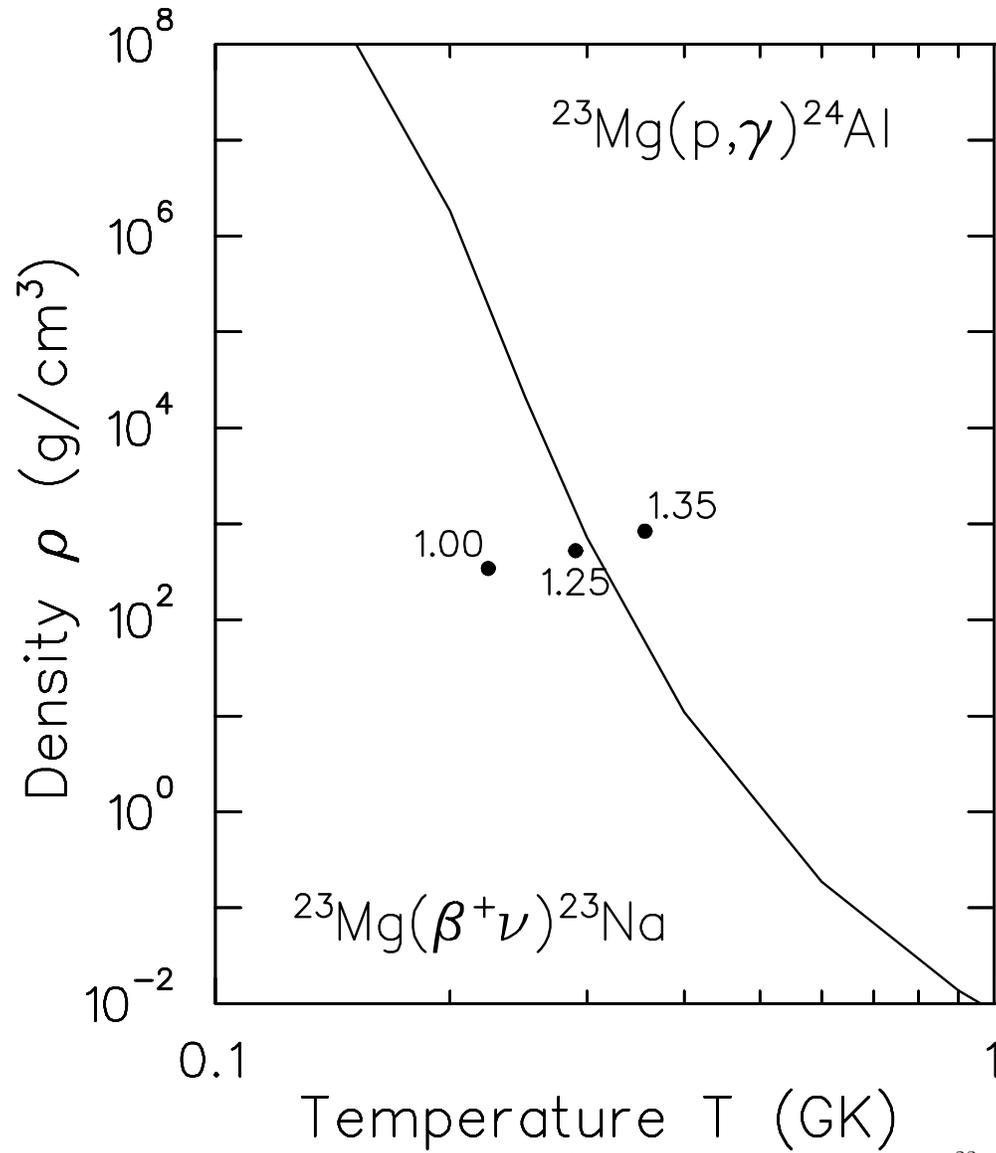,width=13cm}}
\caption[fig3]
{Temperature-density boundary at which the proton capture reaction on $^{23}$Mg
and the $^{23}$Mg $\beta$-decay are of equal strength, assuming a hydrogen mass
fraction of X$_{H}$=0.365 \cite{POL}. Peak temperature and density conditions
achieved in the nova models of Ref. \cite{POL} are indicated by full circles
(see text).} 
\label{fig-taurho}
\end{figure}
\end{document}